# Practical design of alternating-phase-focused linacs

R. A. Jameson

Los Alamos National Laboratory (retired, Guest)
Inst. Angewandte Physik (Guest Professor), Goethe Uni Frankfurt, Max-von-Laue Str. 1, Frankfurt-am-Main, D60438 Germany, email jameson@riken.jp

Conventional magnetic transverse focusing in conventional linacs represents a high fraction of their cost and complexity. Both transverse and longitudinal focusing can be obtained from the rf field by using the strong-focusing effect of alternating patterns (sequences) of gap phases and amplitudes – known as Alternating-Phase-Focusing (APF). Simple schemes have small acceptances and have made APF seem impractical. However, sophisticated schemes have produced short sequence APFs with good acceptances and acceleration rates that are now used in a number of practical applications. Although studied for decades, the design of suitable sequences has been difficult, without direct theoretical support, inhibiting APF adoption. Synthesis of reported details and new physics and technique result in a new, general method for designing practical APF linacs. Very long sequences with high energy gain factors are demonstrated, motivated by the desire to accelerate very cold muons from ~0.340-200MeV (factor 600). An H+ example from 2-1000MeV (factor 500) is also given (the problem scales with the charge-to-mass ratio). The method is demonstrated with simple dynamics and no space charge – incorporation of space-charge and more accurate elements is straight-forward. APF linacs can now be another practical approach in the linac designer's repertoire. APF can be used in addition to conventional magnetic focusing, and could be useful in minimizing the amount of additional magnetic focusing needed to handle a desired amount of beam current.

## I. Introduction – Alternating-Phase-Focusing

APF acceleration has been studied for decades as it offered the possibility to avoid the high cost of magnetic focusing in the linac. Particles exposed to an rf field in a gap may receive focusing or defocusing forces in the transverse and longitudinal directions, according to the phase and amplitude of the gap field. Arranging a sequence of gaps in some particular manner, termed here an "APF sequence", can provide large simultaneous transverse and longitudinal acceptances with high acceleration rates and good emittance preservation, even better in some cases than the well-known RFQ or linacs with magnetic transverse focusing.

Typical transverse focusing and energy gain per unit length (dW/dz) forces over the full 360° range of rf phase (phi) are shown in Fig. 1.

$xvsefoc = \pi qom E_o T sin(phis) \lambda nblt^2/(\beta\gamma^3)$,

$dWdz = E_o qT cos(phis)$

where $qom$ is the charge-to-mass ratio, $E_o$ is the average accelerating gradient, $T$ is the transit-time-factor, $phis$ is the rf synchronous phase at the gap center, $\lambda$ the rf wavelength, $nblt$ is the nominal number of $\beta\lambda$ in one cell (depending on the structure), $\beta$ and $\gamma$ are the relativistic factors.

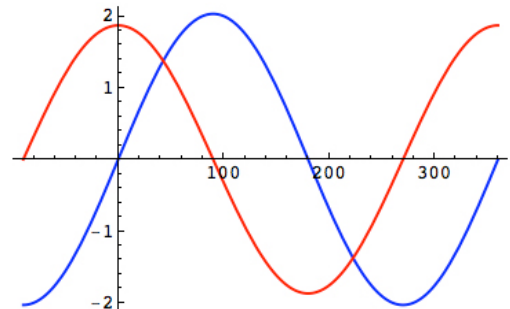

Fig. 1. Typical longitudinal acceleration (red) and transverse focusing forces (blue) over the full 360° range of phi.

There has been no adequate general theory or convenient method for determining an APF sequence with large energy gain. There are some APF linacs in operation, but their designs were laboriously produced by hand, and the sequences are short. Further APF development has also been hindered by a misconception that APF acceptances are necessarily small.

The purpose of this paper is to demonstrate a practical method for designing and optimizing APF linacs. Sec. II. shows the most successful present designs, and summarizes, from the detailed research, important factors needed to synthesize a new general method. Sec. III outlines the method. In Sec. IV, the initial design and optimization steps for a 0.34-20 MeV muon APF linac are outlined, as an example. The development continues in Sec. V using a proton APF linac as the second example, just to demonstrate that the problem scales with charge/mass ratio, and reaches the conclusion that a more sophisticated optimization procedure is required. Sec. VI gives a new optimization procedure based on intelligence derived from APF physics and modern control theory, and applies the procedure to the H+ and muon APF linacs. Next steps for further development are outlined in Sec. VII, and Sec. VIII concludes the development.

## II. The APF sequence, and summarization of important factors

The most sophisticated work was realized in the USSR during the 1960's-1980's [1,2, and culminated in the "Garaschenko Sequence" [3]. Simple schemes using modulation of the synchronous phase by ±Δphis had led to the realization that large swings could work (e.g. ±60°, and extended variations of these schemes, including offsets of the average phis), and it had been realized that smooth modulations over the full range of ±90°, with appropriate variations as the particle velocity increases, were possible.

### A. Garaschenko APF sequence

Garaschenko's 51-cell synchronous phase sequence for a 0.0147-1.0 MeV/u (factor 68), $_{238}U^{7+}$ (qom=1/24), 6.8m, 25 MHz uranium-ion accelerator is shown in Fig. 2. It is seen that it has quasi-sinusoidal features.

It is the result of a complicated nonlinear optimization procedure to find the maximum longitudinal and transverse acceptances. The longitudinal acceptance was larger than for a typical RFQ.

Typically, a nonlinear optimization program must be given good enough starting points that it can converge to the correct optimum. Here the preliminary sequence was based on the extant schemes and a successfully operating APF linac at Dubna, which set out the general properties of the sequence in six separate focusing periods of 6-13 gaps and different spacings in each.

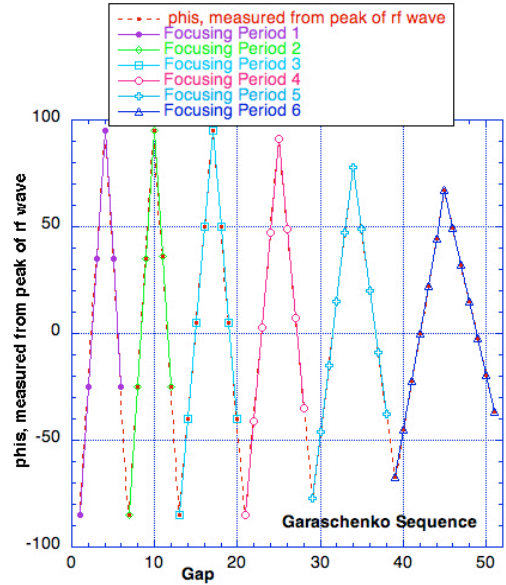

*Fig. 2. Garaschenko APF sequence. Phis is measured from the peak of the wave.*

The great value of this paper is not that it helps find a good initial sequence, but that *it shows the form of an optimized sequence*. From this, we can draw very useful general conclusions.

## B. NIRS APF sequence

The IH linacs now in operation at NIRS, U. Gunma, and similar therapy machines under construction in Japan [4], also have an APF synchronous phase pattern very similar to that of Garaschenko. A general sequence function for the $C_4^+$ (qom=1/3), 0.608-4.0 MeV/u (factor ~6), 3.44m, 200 MHz NIRS linac was written as:

$$f_s(n) = f_0 \exp(-an)\sin((n-n_0)/(b\exp(cn)))$$
where $n$ is the cell number.

The five parameters were searched, and the sequence then optimized for small output energy spread and output matching to the following section. The 72-cell sequence, closely resembling Garaschenko's, is:

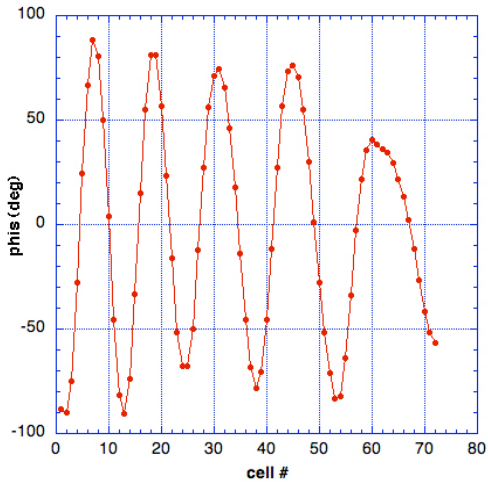

Fig. 3. The NIRS APF sequence. (Courtesy of NIRS.)

A similar formula (see Eqn. 1 below) can be fitted to the Garaschenko sequence.

## C. Collection of important APF research details

Overall, many different sequence schemes have been developed, and important details discovered. Summarizing these leads to a practical design method.

In 1988 during research at KEK, the author explored the effect of superimposing various field and synchronous phase error patterns on a long 1 GeV magnetically-focused proton linac for waste transmutation, and found that certain patterns of sinusoidal modulation produced an APF, or anti-APF effect. A generalized error function with a number of parameters was used, and searches made over the parameter space.

The author had gathered a large collection of APF papers (not so easy at that time), and during the second half of that year at the Keage Accelerator Institute of Kyoto University, under the leadership of Prof. H. Takekoshi, his student H. Okamoto became interested in APF and used the smooth approximation method to characterize the stability region, acceptances, and other features of the sequence [5]. This well-known method leads to the Mathieu-Hill equations, for which suitable sequence trajectories can be investigated on a transverse stability chart. The method is useful for understanding the general properties of APF sequences but limited because of the assumptions of periodicity, small phase advances, and no acceleration and thus is not well suited to determining an actual sequence with acceleration. However, as shown below, it has an important utility in the synthesis of the new method.

During the next decade, work in Russia continued, especially by V.V. Kushin, V.K. Baev, and S. Minaev, who was a leading practitioner of actual APF designs until his death in 2010 and who influenced many extent APF designs.

However, Minaev noted in [6] that "there is no theory for optimization of drift tube array so far", and this is still the situation at this date. In this paper, Minaev shows that he uses a quasi-sinusoidal sequence progression similar to that shown by Garaschenko, where the period of the sinusoid lengthens as the beam accelerates, but the three succeeding sections are optimized in turn by hand.

The difficulty of deriving a sequence is a main reason why previous APF designs have been for short sequences. Another is the misconception that APF acceptances are necessarily small. However, even the short sequence examples above show the possibility of high energy gain per meter, and high energy gain ratios (x68 for the Dubna linac). A motivation for this paper is to facilitate design and optimization of long sequences with high energy gain ratios.

The extensive literature contains details that are important for the synthesis of a new method.

Different spatial harmonics of an accelerating structure can be used for the transverse focusing and the longitudinal accelerations [7,8]. The latter paper also points out that in the Alvarez $E_{010}$-mode structure, the stored energy is distributed uniformly, and therefore the rf voltage applied at each gap is automatically proportional to the cell length, which is advantageous for APF design and practice. It was shown [9] that alternating both the field phase and amplitude allows small transverse emittance growth by aligning the sequence more along a line of constant phase advance on the stability chart.

It is noted in [9] and also in [3] that there should be an offset of 5°-10° in the initial average value of the synchronous phase, and [3] also notes that it is helpful to also have a tilt, so that the average synchronous phase tends from ~+5° to ~-5°. From Fig. 1, it is seen that such adjustment will increase the acceptance in one plane, while decreasing it in the other.

As in conventional linacs, as acceleration occurs and the velocity beta increases, the focusing strength can be maintained by lengthening the focusing period.

As seen in Figs. 1-3, the full range of longitudinal focusing is being exploited by the Garaschenko and NIRS sequences. An APF linac employing the range -90° < phis < +90° accelerates efficiently and is similar to a conventional drift-tube-linac (DTL) in length.

Okamoto [10] adds valuable insight into the value of a tilt by pointing out that it helps avoid the danger of emittance equipartitioning [11] via a synchrobetatron coupling resonance.

APF linacs in operation so far are apparently for small beam currents, for which there are many applications. It is of interest to explore APF capabilities for higher beam power as well, because of the potential for large cost savings. Acceptances with space charge will be within the bounds of good small-(zero)-current transverse and longitudinal acceptances, so design at zero current should be done first, and will be faster to execute. This paper concentrates on zero-current design; other effects, such as space charge or combination with magnetic focusing, are straight-forward to add later.

The smooth approximation theory has been extended to include acceleration and amplitude modulation [12, 13], both again using a sequence of sinusoidal form. The first uses six parameters to describe the sequence: equilibrium synchronous phase, phase modulation amplitude, length of APF period, incremental energy gain, plus two additional parameters to include simultaneous modulation of the accelerating field amplitude - the relative phase between the amplitude and phase modulation and magnitude of the amplitude modulation. Stability boundaries are shown, but practical examples are not explored. The latter gives an example for an a superconducting accelerator structure period of one solenoid and two spoke-resonator cavities, one focusing and the other defocusing, demonstrating good transmission of up to 50 mA over an energy gain of ~x4.

### D. Sequence optimization

However, it is apparent in Figs. 2 & 3 that the optimized sequences are only quasi-sinusoidal, and therefore a method for determining a preliminary sequence is needed, to be followed a procedure for subsequent optimization by a typical nonlinear constrained optimization program for required characteristics like maximum transmission, smallest energy spread, minimum emittance growth, etc. The significant deviations from the formulas in the Garaschenko and NIRS sequences are the result of careful final optimizations, using a nonlinear optimization technique by Garaschenko, and a Monte Carlo method by Tsutsui at NIRS.

A similar method is outlined in [14] for design and optimization of a short 26-cell sequence, which resembles Figs. 2 & 3, to accelerate $_{238}U^{34+}$ (qom=1/7) from 0.140 MeV/u to 0.542 MeV/u (factor~4) in 2m (~2.6MV/m) at 52 MHz. The optimization minimizes a desired objective function by adjusting the 26 gap phases directly.

Using the gap phases as the optimization parameters directly can work for short sequences, but even then may have difficulty converging. If the gap amplitudes are to be optimized as well, the number of variables doubles, and for long sequences, the difficulty compounds further. Therefore another strong motivation for this paper is to provide more information to the optimization procedure for quicker and more accurate convergence.

### III. Synthesis of a practical APF design, simulation, and optimization method

The present situation is that practical methods have seemed cumbersome and limited to short sequences. As there seems to be no inherent limitation to the beam energies for which APF can be applied, it is of interest to synthesize a general method to explore long sequences with large energy gain, and efficient optimization. If reasonable zero-current transverse and longitudinal acceptances can be found, then APF-focused beams with space-charge up to some limit will also be possible.

By considering the general principles underlying the details of a large amount of previous work as outlined above, a generally practical and straight-forward method is synthesized for generating and optimizing APF linacs, as follows:

1. An optimal sequence can be based on a general sinusoidal form for modulation of the gap synchronous phase *phis*, with a 7-parameter function of the cell number *ncell* (*radian (=π/180°) indicates that the corresponding parameters are in degrees):

```
phis =  phioffset*radian -
phitilt*radian*ncell +
phiampl*radian*Exp[-
phiatten*ncell]*Sin[2.*Pi*ncell/
(phiperiod*Exp[peratten*ncell]) +
phistart*radian]           (1)
```

As indicated above, it may be desired for further reduction of transverse emittance growth to apply sequences to both the gap phase and amplitude. The method can be extended for this. The form of the attenuations can be changed. The APF linac is then designed by simulation.

2. A 7-dimensional grid search over the seven parameters can be performed quickly for zero beam current with a fast multi-particle simulation code, with finer and finer grids.

3. When parameters are found which give an initial adequate transmission, the sequence is optimized, using a new strategy, for the desired beam properties using nonlinear optimization techniques.

4. The modeling can then be refined, with more accurate modeling, with space-charge, etc., and the process repeated.

### *A. LINACSapf dynamics method*

A new simulation code, LINACSapf, was written for APF linac design and simulation.

A simple chain matrix representing each cell as a drift and an rf gap can be used. The linac is designed sequentially cell by cell, and thus acceleration is taken into account in fitting the phase sequence. The cell lengths are irregular, as determined by the local phase difference across the cell from the APF phase sequence, and the local velocity β. The Trace3D code rf gap transformation is used at the end of the cell.

The method is also well-suited for simulation of sequences with more accurate gap models (e.g. also including underlying magnetic focusing), and for beams with space-charge.

It is useful for preliminary design work not to insist on completely realistic conditions. The gap voltage should be realistic, as determined from the Kilpatrick Criterion and structure peak field characteristics. However, at first the aperture should not be a restriction. Zero beam current and small input beam emittances are useful when searching for workable sequences. In this paper, the beams are input with upright ellipses ($\alpha$'s = 0). The program can easily compute the acceptances, and this matching information can be used at any time to correct the input distribution.

The linac generation and dynamics are embedded in a driver, and execution is very fast, enabling parameter scans with thousands of runs to be performed quickly.

The beam performance is sensitive to the APF sequence parameters, and this made the initial programming for setting up various traps and debugging difficult *.

## IV. Initial design of a 0.34-20 MeV muon APF linac

The KEK Muon Project [15] requires acceleration of a small number of very cold muons (qom = 1/ 0.11272122345 = 8.87) over the range of ~0.34-200 MeV, with small transverse emittance growth and minimal output energy spread. The proposed muon beam current is very small and space charge can be neglected. The APF linac is an attractive candidate, because significantly lower cost is realized when magnetic focusing elements are not required.

RFQ designs to bring the muon beam to 0.340 MeV have been developed [16]. Case 1.3 of that study has an output beam with transverse normalized rms emittance = 0.32 mm.mrad, a phase spread of ±22°, and an energy spread of ±2.1%. This beam is input to a 324 MHz APF linac, for which 1.8*(Kilpatrick limit)=0.32 MV/cm. A flat field and a peak-to-gap field ratio of ~6 are used to determine the working $E_o$ ~ 0.055 MV/cm. The initial bore radius is 0.5 cm, with linear expansion to ~1.5 cm after the ~23 cells (~6.5 m) required to reach a final energy of ~22 MeV, where $\beta$=~0.56. Particles are considered "accelerated" if their final energy is within 3% of the final synchronous energy.

For the first stage of the study, a design for 0.34-20 MeV is sought. Only ~23 cells are required – a short sequence, for which the simulation is extremely fast.

### A. Initial parameter search

Fig. 4 shows three successive searches over the six parameters of Eqn.(1) with *phiampl* = 90°.. The first search showed a tendency toward an optimum transmission, which was then found in the second search with narrower parameter ranges.

The case for best transmission showed transmission of 99.75%, but accelerated fraction of only 67.3%. The parameters are shown in Table 1.

---

\* For example, the Trace3D gap transformation starts with the standard formula for the new energy after the gap as the previous cell's synchronous energy plus the energy gain across the gap, W(n+1) = Ws(n) + qEoTLcos(phi), where Ws(n) is the present cell's synchronous energy, q is charge, Eo is electric field, T is transit-time-factor, L is cell length, phi is the rf phase at the gap center. The gap transformation then finds the new energy again as the new cell's synchronous energy Ws(n+1)plus an energy offset $\Delta$w. The two new energies are usually very close, but for APF sequences with bad performance, there could be significant difference, which contributed to initial confusion. With good APF sequences, the difference is very small – the Trace3D transformation is correct.

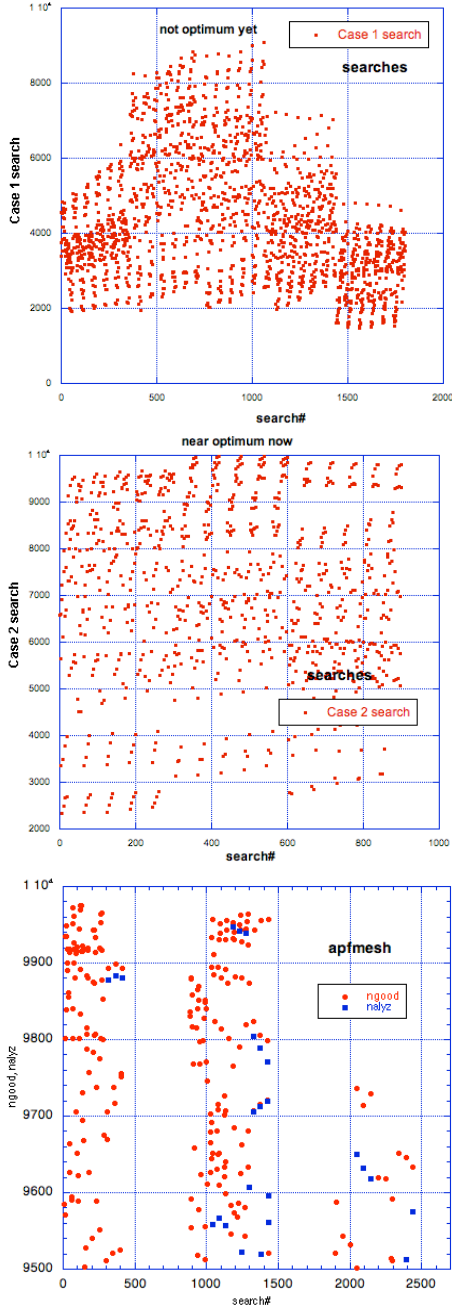

The third search looked for best accelerated fraction; the third graph in Fig. 4 shows that this is reached at different parameters. The transmission is 99.53%, with accelerated fraction 99.47%, and the parameters are shown in Table 2:

| phioffset | 25.0 |
| --- | --- |
| phitilt | 1.00 |
| phiampl | 90.0 |
| phiatten | 0.008 |
| phiperiod | 2.3 |
| peratten | 0.009 |
| phistart | 55.0 |

Table 2. Initial parameter search result for best accelerated fraction.

Fig. 5 shows the two phase sequences, and it is seen that they are quite different. It is interesting to note that the tilt parameter is invoked for best accelerated fraction.

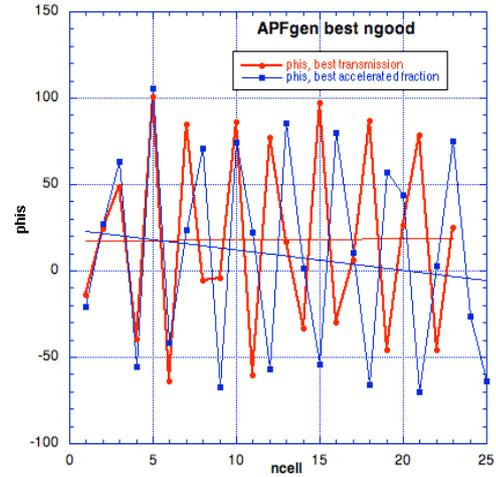

*Fig. 4. Results of three successive searches over the six parameters of Eqn.(1). The vertical axis is the number of particles transmitted (ngood – red) and the number of particles accelerated to within 3% of the final synchronous energy (nalyz – blue).*

*Fig. 5. The APF gap phase sequences for the best transmission (red) and best accelerated fraction (blue) cases. The linear fits show the effect of the tilt parameter.*

| phioffset | 20.0 |
| --- | --- |
| phitilt | 0.00 |
| phiampl | 90.0 |
| phiatten | 0.01 |
| phiperiod | 2.3 |
| peratten | 0.008 |
| phistart | 50.0 |

Table 1. Initial parameter search result for best transmission.

The output x-xp and dphi-dw phase spaces for the two cases are shown in Fig. 6. As there is zero current, the beam remains a filament, as evidenced in the longitudinal phase space. Although almost all of the input muon beam has been captured and accelerated, smaller final transverse emittance and energy spread are to be sought.

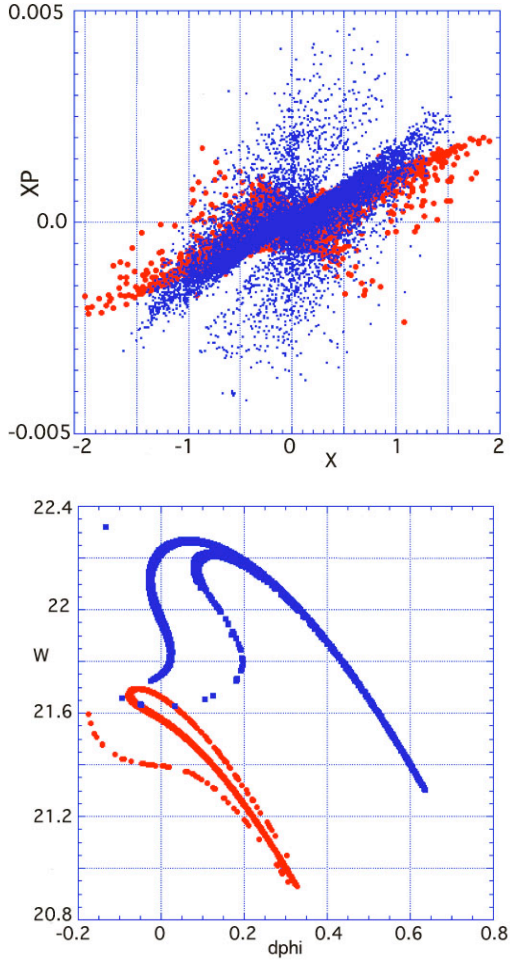

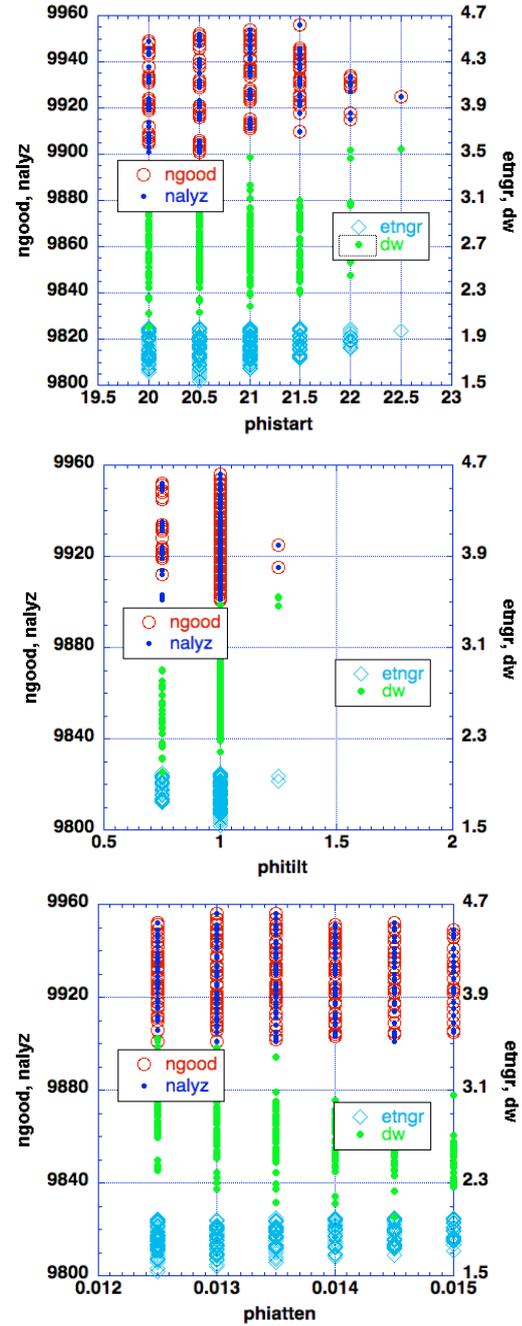

*Fig. 6. Above: x-xp (red), y-yp blue)). Below: dphi-W phases spaces for the best transmission case (red) and the best accelerated fraction case (blue).*

## B. Find minimum output transverse emittance and energy spread

The output transverse emittance and energy spread are computed and the smallest emittance growth and smallest energy spread (%) are sought.

Fig. 7 shows search results on finer meshes. 10000 particles were computed, and the run kept only if the number of accelerated particles was ≥ 9900 (99%). For each of the six varied parameters, the number of transmitted particles (ngood), the number of particles accelerated to within 97% of the final synchronous energy (nalyz), the transverse normalized rms emittance growth (etngr), and the energy spread (dw,%) are shown, giving information on the optimum and its sensitivity.

The last graph shows a correlation between the transverse rms emittance growth and the output energy spread. Lower energy spread is obtained at the expense of more transverse emittance growth.

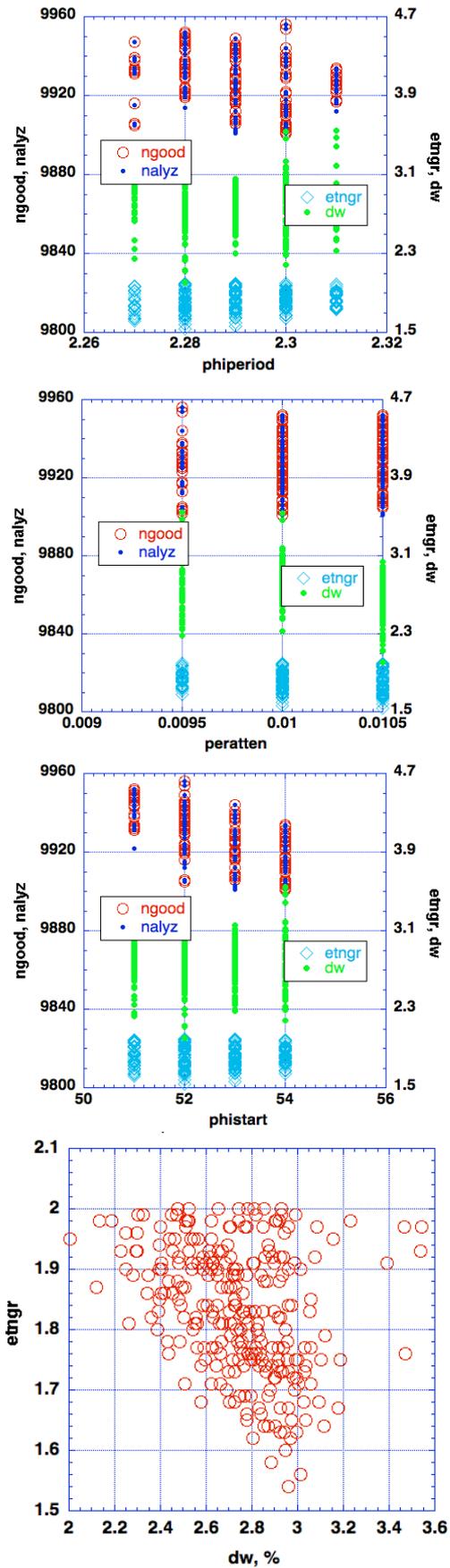

*Fig. 7. Results of parameter search on finer grids.*

The parameters and output characteristics which gave minimum energy spread (Case 430), are shown in Table 3. and in Fig. 8:

| phisoffset | 10.0 | #cells | 23 |
|---|---|---|---|
| phitilt | 0.75 | length,m | 7.14 |
| phiampl | 90.0 | Xmsn,% | 99.24 |
| phiatten | 0.0145 | Accel,% | 99.24 |
| phiperiod | 2.28 | etn growth | 1.95 |
| peratten | 0.0105 | output energy spread,% | 2.0 |
| phistart | 52.0 | | |

Table 3. Parameters and output characteristics for minimum energy spread.

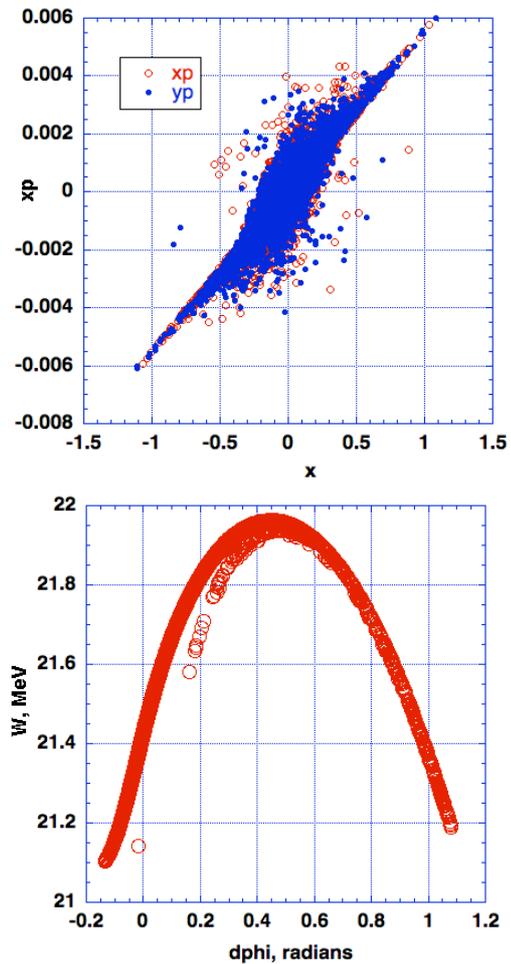

*Fig. 8. Output x-xp, y-yp and dphi-W phase spaces for minimum energy spread case.*

The parameters and output characteristics which gave minimum transverse emittance growth (Case 2926) are shown in Table 4. and in Fig. 9:

| phisoffset | 20.5 | #cells | 23 |
|---|---|---|---|
| phitilt | 1.0 | length,m | 7.05 |
| phiampl | 90.0 | Xmsn,% | 99.29 |
| phiatten | 0.0125 | Accel,% | 99.29 |
| phiperiod | 2.28 | etn growth | 1.54 |
| peratten | 0.0105 | output energy spread,% | 2.96 |
| phistart | 52.0 | | |

Table 4. Parameters and output characteristics for minimum transverse rms emittance growth.

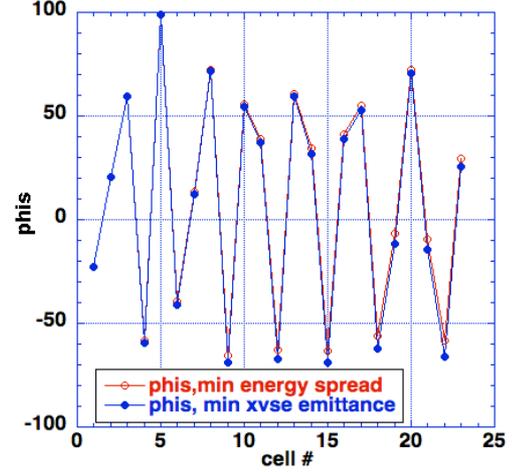

Fig. 10. The phis sequences for the minimum energy spread and the minimum transverse normalized rms emittance cases.

Fig. 10 shows the differences between the phis sequences for these two cases.

### C. Search using a constrained nonlinear optimization program.

*1. Optimization on the phis's of the sequence*

The individual phis values at the 23 cells (23 variables) are optimized by the constrained nonlinear optimization solver NPSOL [17], for various objective functions:

    1. Best transmission
    2. Best accelerated beam fraction
    3. Minimum transverse normalized emittance (etn) growth
    4. Minimum output energy spread (dw), %
    5. Minimize the sum of etn growth and dw/2. dw is divided by 2 for good weighting against the etn growth.
    6. Minimize the sum of etn growth, dw/2, and (npoints - accelerated points)/20. npoints = 10000, and the divide by 20 is for proper weighting.

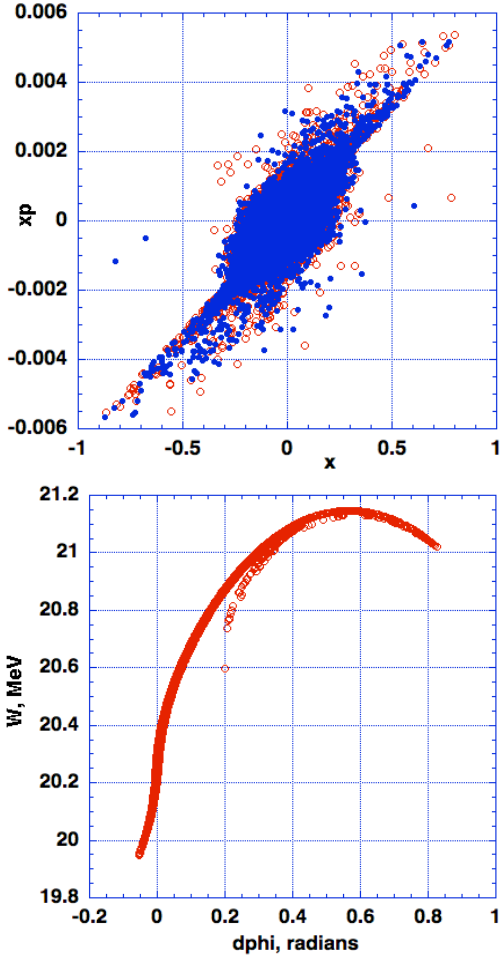

Fig. 9. Output x-xp, y-yp and dphi-W phase spaces for minimum transverse rms emittance growth case.

The difference in energy spread is a bucket rotation effect.

As is generally the case with nonlinear optimizers, settings for the constrained range on each of the working variables, and on various features of the optimizer such as the way it searches for and handles derivatives, must be explored. Also the starting solution must be close enough to the final solution that the optimizer does not converge to a false local minimum. In this case, bounds of ±5° to ±10° on the phases were used, along with appropriate settings for the difference intervals.

A starting sequence was selected midway along the lower correlation boundary of the last graph in Fig. 7. As would be expected, the initial fine mesh search gives a good starting sequence, and not much improvement is found by the optimization, but it does succeed in finding the desired improvements (Table 5).

| Objective function | Xmsn, Accel, % | etn growth | energy spread |
|---|---|---|---|
| Starting sequence | 99.06 | 1.683 | 2.681 |
| 1. max xmsn | 99.27 | 3.534 | 2.773 |
| 2. max accelerated | 99.40 | 2.187 | 2.916 |
| 3. min etn growth | 95.05 | 1.674 | 2.679 |
| 4. min energy spread dW, % | 95.39 | 2.137 | 1.566 |
| 5. min (etngrowth, dW/2) | 99.18 | 1.230 | 1.714 |
| 6. min (etngrowth, dW/2,(npoints-naccel)/20) | 99.28 | 1.401 | 1.988 |

Table 5. Results of optimization of the 23 phis points directly.

### 2. Comparison of starting sequence and optimized sequence

The starting sequence was generated by the general 7-parameter sequence formula (1). The parameters of the initial (starting) sequence) were:
```
phisapf /. {phioffset->20.5,
phitilt->1.00, phiampl->90.0,
phiatten->0.0145, phiperiod->2.28,
peratten->.0105, phistart->52.}
```

Mathematica FindFit can fit the parameters to the optimized sequence. The unconstrained natural form for Case 6 in Table 1 gives:

```
Gfit = FindFit[sequence,
 phisapf, {{phioffset,20.5},
 {phitilt,1.00},{phiatten,0.0145},
 {phiperiod,2.28},{peratten,.0105},
 {phistart,52.}},ncell]
```
where the starting values are those of the parameters of the starting sequence. The result is:

```
{phioffset->17.0621, phitilt->
0.685082, phiatten->0.0161262,
phiperiod->2.28007, peratten ->
0.0105318, phistart->52.2236}
```

The differences from the initial sequence are shown in Fig. 11:

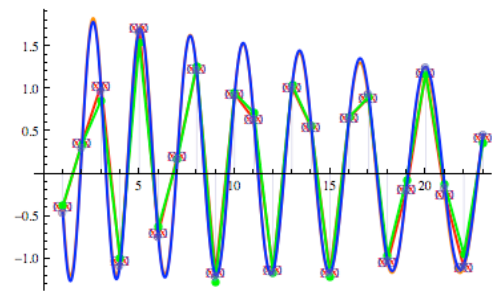

Fig. 11. The starting sequence curve from the general 7-parameter sequence formula (orange), and the synchronous phases at the 23 cell gaps (red). The ordinate is radians. Plotted over these are the optimized sequence curve from FindFit to the general 7-parameter sequence formula (blue), and the synchronous phases at the 23 cell gaps (green).

### 3. Optimization on the seven parameters of the sequence

Optimization on the seven parameters of the sequence was rather sensitive to the bounds and the optimizer settings, probably because there is less information to work with. Optimization on the gaps is recommended.

### 4. Optimization on the phis's and the (gaplength/bl)'s "gapobl" of the sequence

It was noted above that a tip in the references indicated that less transverse rms emittance growth might be obtained by varying both the synchronous phase and the accelerating field amplitude at the sequence gaps.

The energy gain per cell = VTcos(phis), where the transit-time-factor T = sin(gap length/βλ)/ (gap length/βλ) = sin(gapobl)/gapobl. The Trace3D convention is that "gap voltage" = V = E0*cell length because E0 is integrated over the cell length. The cell lengths are already determined from the phis sequence.

As no information exists on the possible or optimum form of a gapobl sequence, whether in conjunction with a phis sequence or not, the constrained nonlinear optimization procedure is very useful.

Optimization was performed on the (gap length/βλ) and phis at each cell, with ±5° bounds on phis and 0.13 – 0.25 on gapobl. The starting sequence is the result of Table 5 Case 6. The objective function was set as a weighted combination of the emittance growth, the output energy spread, and desiring a large accelerated fraction:

```
objfunc=2.d0*etngrowth+energysprd+
dble(npoint-nalyz)/100.d0
if(objfunc.lt.0.d0)objfunc=1000.d0
```

The results: 97.31% accelerated, etn growth = 1.314, output energy spread 1.016%.

In comparison with Table 5 Case 6, there is less accelerated fraction, but adjusting also the gap lengths appears to indeed produce somewhat less transverse emittance growth and also less output energy spread.

The sequences are shown in Fig. 12.

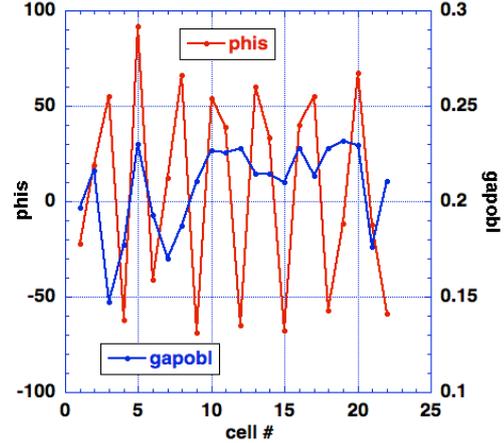

*Fig. 12. Sequences resulting from optimization on the phis's and the gap lengths.*

The output phase-spaces are shown in Fig. 13.

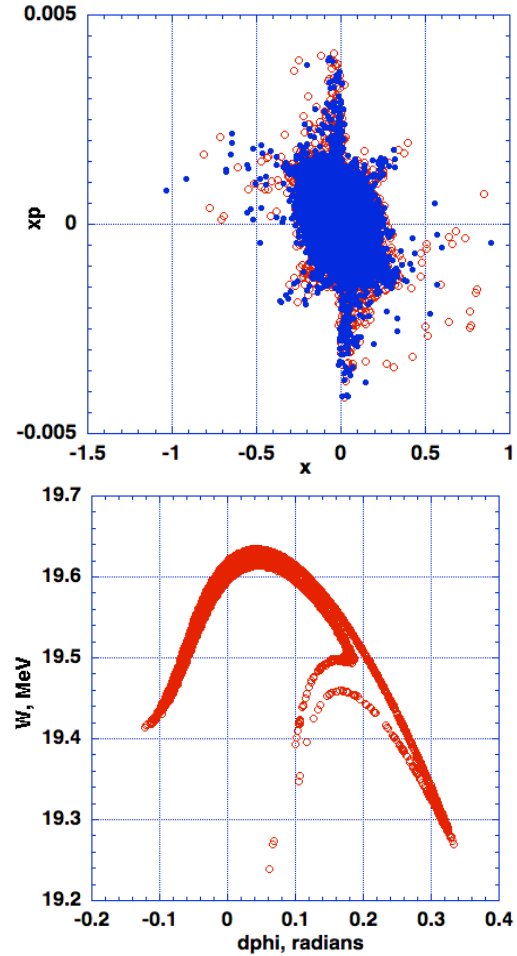

*Fig. 13. Output x-xp, y-yp, and dphi-W phase space for optimization of both the phis's and the gapobl's.*

The other objective functions used in Table 5 did not work well, tending to much reduced accelerated fraction.

*(e) Reduce the aperture expansion, try other apertures*

As stated, the APF linac above had initial aperture = 0.5 cm radius, and final aperture = ~1.5cm radius.

At different aperture expansions, the beam loss changes (Table 6):

| Final Aperture | Accel fraction, % | etn growth | dw |
|---|---|---|---|
| 2 | 99.22 | 1.344 | 1.016 |
| 1.5 | 97.31 | 1.314 | 1.016 |
| 1 | 89.77 | 1.164 | 1.016 |
| 0.5 | 69.25 | 0.856 | 1.016 |

Table 6. Case optimized on both phis's and gapobl's. Initial aperture is 0.5 cm. Effect of linear aperture expansion.

Reoptimizing with constant aperture = 0.5 gave 76.66% accelerated, etn growth = 1.381, dw = 0.851.

The rms transverse emittance rms ellipse (betax+betay)/200 was added to the optimization function, yielding very small improvement in the accelerating fraction.

The beam loss occurs early with 0.5 cm initial aperture and 1.5 cm final aperture. Opening the initial aperture and keeping it constant gives good accelerated fraction (Table 7).

| Aper, cm | Aper coeff | Accel frac, % | etn growth | dw |
|---|---|---|---|---|
| 0.75 | 0.0 | 92.77 | 1.260 | 1.074 |
| 1.0 | 0.0 | 99.38 | 1.436 | 1.074 |
| 1.5 | 0.0 | 100.00 | 1.490 | 1.074 |

Table 7. Beam loss vs. cell number for 0.5 cm initial aperture and 1.5 cm final aperture. The table shows improvement of accelerated beam fraction with larger, constant apertures.

## V. Initial result for 2-200MeV H+ APF linac

The research then proceeded with protons, just to illustrate the same procedure with a different particle (the problem scales with the charge-to-mass ratio). A good design for H+ at 324 MHz from 2-20MeV similar to the above muon design was quickly found, so a longer sequence for 2-200 MeV (factor 100) was explored next.

It was learned that the mesh search over the sequence parameters must be done with both fine enough resolution and with wide enough range, because there can be several areas of parameter space where a significant percentage of accelerated particles is obtained, but one area will be best. It is useful to plot the formula. The phis sequence swing should be kept large over the desired energy range to maximize the acceleration rate. The period attenuation is sensitive and if too large can overwhelm the sinusoid.

For easier control of the attenuation, the sequence formula was changed to linear attenuation of the period:

```
phis =  phioffset*radian -
phitilt*radian*ncell +
phiampl*radian*Exp(phiatten*ncell)*
Sin[2.*Pi*ncell/
(phiperiod*(1.d0-peratten*ncell)) +
phistart*radian]          (2)
```

~164 cells are required, with total length ~45 m. The best mesh search result was with parameters:
```
{phioffset->22.°, phitilt->0.2,
phiampl->90°, phiatten->0.006,
phiperiod->4.4, peratten->0.005,
phistart->56°},
```
with transmission of only 39.23% and accelerated fraction of 36.47%.

Now optimization to raise the accelerated fraction became difficult.

Several methods were ineffective and difficult to make converge:
   - optimization on all sequence points directly.
   - optimize successively each sequence point individually.
   - optimize from each cell to the end successively.
In each case, the objective function is computed over the whole linac.

It was necessary to find a better optimization strategy.

### VI. Intelligence for a new optimization procedure from APF physics and modern control theory

The optimization strategy from a given local point to the end successively is based on the principle from modern control theory (and for any journey) that it is most effective to reoptimize from each present state to the end. Even so, enough information must be available on sensible paths for convergence; otherwise a solution may not be found, or found only with a large amount of unnecessary work. As indicated above, adjusting the whole matrix of remaining cell phases and gap lengths may eventually converge with an extreme amount of computation, or may not[1].

Applying this principle, new information is synthesized from the detailed APF development outlined in Sec. II.

From above, recall that the Garaschenko [3] and Minaev [6] sequences use adjoining sequences of varying period length (N); Garaschenko uses six sequences with N = {6,6,8,8,10,13}. The NIRS or Eqn. (1 or 2) sequences use a sinusoid with period `(phiperiod*(period attenuation function))`, where the attenuation function is of some form like `Exp[peratten*ncell]` or `(1-peratten*ncell)`.

The local N at each cell is obtained by applying the sequence formula (1) or (2) at that cell, and computing the ncell ahead for which the period accumulates by 1. (phase advance of 2*π).

The new optimization strategy is then, at each cell sequentially, to optimize over the local period length N, with the objective function computed for the whole linac.

The transverse and longitudinal phase advances from the smooth approximation for the focusing characteristics of a beamline contain information about both the external fields of the linac or beam line, and also about the beam itself, in terms of the space-charge forces or other effects. When the parameters of the linac or beam line change slowly with respect to the betatron (transverse) or synchrotron (longitudinal) oscillation wavelengths of the phase advances, the smooth approximation formulae can be applied locally at each cell, and are in principle sufficient for "beam-based" ("inside-out") design of any linac [18], although not, as seen here, for specifically optimized APF.

In the cited references, the smooth approximation phase advances are derived for a general APF sequence. Instead, in LINACSapf, the zero-current transverse and longitudinal phase advances s0t and s0l are computed locally at each cell from the rms envelope equations. (These quantities are often negative, so the absolute value is used to present the result in interpretive form.).

---

[11] This is directly analogous to the difficulties existing in correcting beam orbits, where the "experts" gathered data from all instrumentation (such as beam position monitors) at once and attempted to find correcting information to the beam control devices (such as beam steerers) by inverting the data matrix. This resulted in corrections to all of the control devices. In contrast, in the early-1980's the subject of beam-based control was emerging, and new tools were contemplated from new fields such as "artificial intelligence (AI)" and "expert systems". I dispatched S.H. Clearwater from the Los Alamos AT-Division as a Post-Doc to SLAC to work with M.H. Lee, and a full system of programs (COMFORT, ABLE) was developed and implemented on the SLC control system, which could efficiently identify where a beam orbit error, or beam instrumentation error, actually occurred and then made local corrections. (The experienced reader will wonder how it turned out, and will with understanding nod to learn that the "matrix-inversion" school nevertheless won, and the AI work was forgotten…)
"Prototype Development of a Beam Line Expert System", S.H. Clearwater & M.J. Lee, 1987 PAC, p. 532; "Error-Finding and Error-Correcting Methods for the Startup of the SLC", M.J. Lee, S.H. Clearwater, S.D. Kleban, L.J. Selig, 1987 PAC, p. 1334; "Modern Approaches to Accelerator Simulation and On-Line Control", M. Lee, S. Clearwater, V. Paxson, E. Theil, 1987 PAC, p. 611.

## A. Results for H+ APF linacs, continued

Optimizing the gap synchronous phases (phis's) over each N for maximum accelerated fraction, resulted in immediate dramatic improvement of transmission and accelerated fraction to ~85%, transverse emittance (etn) growth 2.7, output energy spread (dw) ~±1.0%

Further experimentation with the bounds on the phis's and gapobl's, and some iteration, resulted in transmission and accelerated fractions of >94%, etn growth ~2, dw <±1% (aperture 1.5-3.0 cm)

The changes to the H+ 324MHz 2-200MeV original sequence (from the mesh search) by optimization on phis only with ±5° bounds are shown in Fig. 14.

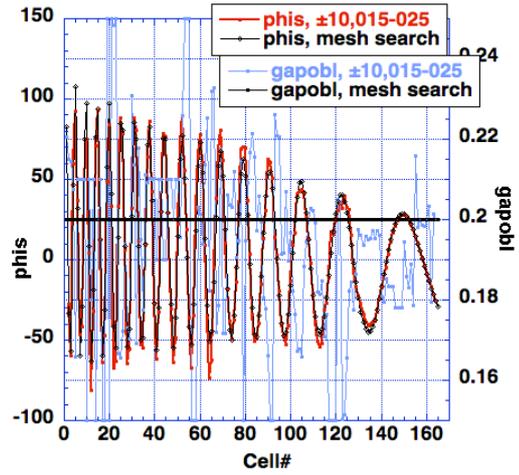

*Fig. 15. Change to original sequence by optimization on both phis with ±10° bounds and on gapobl with bounds 0.15-0.25*

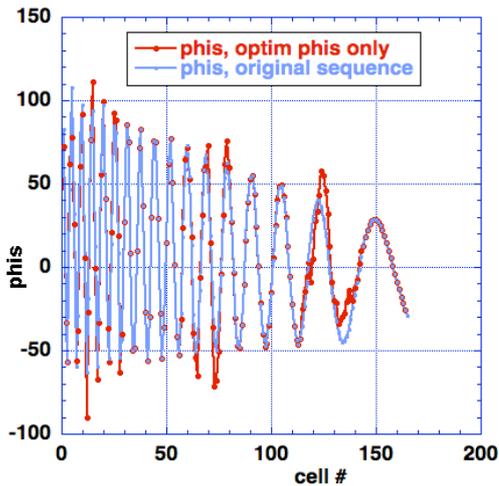

*Fig. 14. Changes to original sequence by optimization on phis only, ±5° bounds.*

Fig. 15 shows the changes to original sequence by optimization on both phis with ±10° bounds and on gapobl with bounds 0.15-0.25.

Random errors in the phis's up to ±2% had only small effect, and even improved the results for some random number sets.

The real, rms transverse acceptance is 2 $\pi$.cm.rad (Fig. 16a). The longitudinal phase capture measured at the upper cusp of Fig. 16b is ~±17° with rms value ~8°, and energy capture +0.010 to -0.045 MeV.

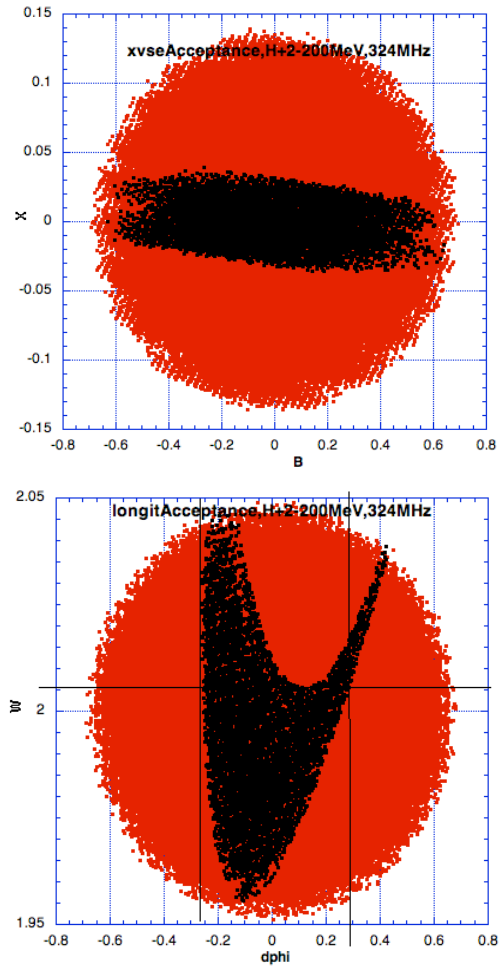

*Fig. 16. a. Transverse acceptance of a 2-200MeV, 324 MHz H+ APF linac. b. Longitudinal acceptance.*

A 200-1000MeV, 972MHz APF linac was then attempted, with the same Kilpatrick factor of 1.6, being careful to keep phis swings high, and not letting the attenuation factor overwhelm the sequence. The mesh search resulted in an input N that matched the output N of the 2-200 MeV linac. The total length is ~174 m, with 771 cells. With no aperture restriction and tiny ±22°, ±0.007MeV bunch. the mesh search yielded a starting sequence and optimization resulted in ~94% accelerated fraction. Acceptance matching gave improvement to >99%, with longitudinal acceptance region ~±20°, ±0.1MeV, Fig. 17.

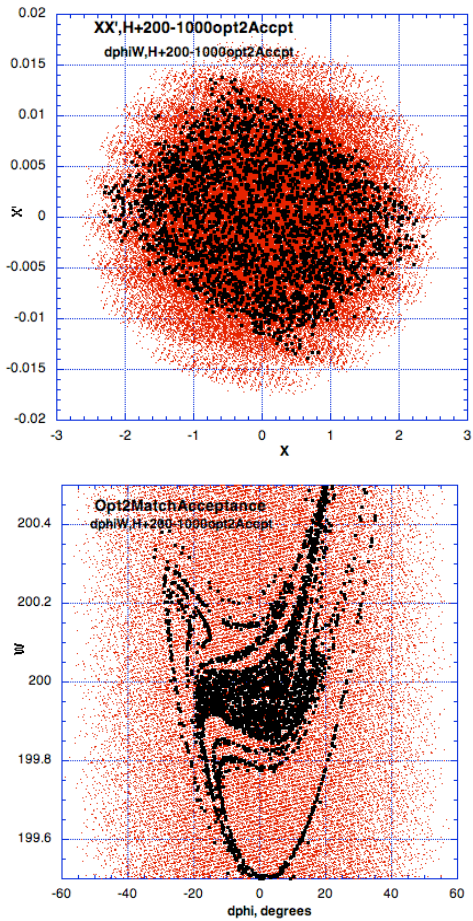

Fig. 17. a. Transverse acceptance of a 200-1000MeV, 972 MHz H+ APF linac. b. Longitudinal acceptance.

A reality check with a ±72° (at 972 MHz), ±1.9MeV beam output from the H+ 2-200MeV design yielded only 8% accelerated.

Some further sequence adjustments or reality factors might raise or lower the acceptances. The acceptance with beam current will be less, but some amount of beam current could be accelerated. *The point is that the design and optimization procedure works even for long sequences and large energy gain factors.*

### B. Results for Muon APF Linacs, continued

Optimizing the 0.340-20MeV, 324 MHZ, 0.5 cm initial aperture and 1.5 cm final aperture muon linac design with the period N procedure on the phis's resulted in about the same performance, ~98% transmission, 98% accelerated, as the direct optimization of the 23 phases. The transverse and longitudinal gap impulses are opposite in sign and the longitudinal impulse is stronger. The phase advances s0t and s0l, Fig. 18a., show that s0t lies essentially under 90° as necessary to avoid the resonance. s0l is above 90° initially and then falls below; this may have a negative effect on the acceptances and needs to be investigated.

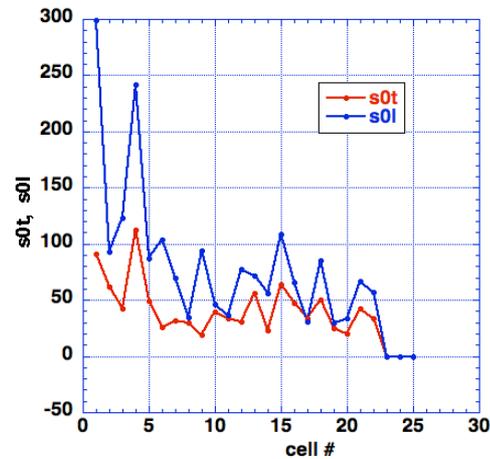

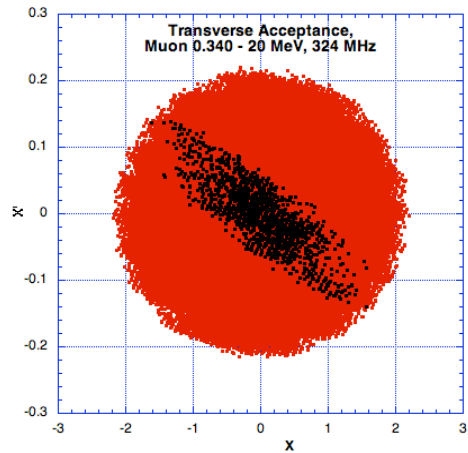

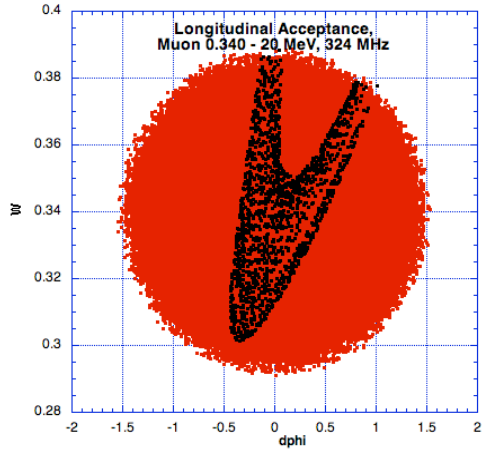

*Fig. 18. a. Zero-current phase advances for muon 0.340-20MeV, 324MHz APF linac. b. & c. Transverse and longitudinal acceptances.*

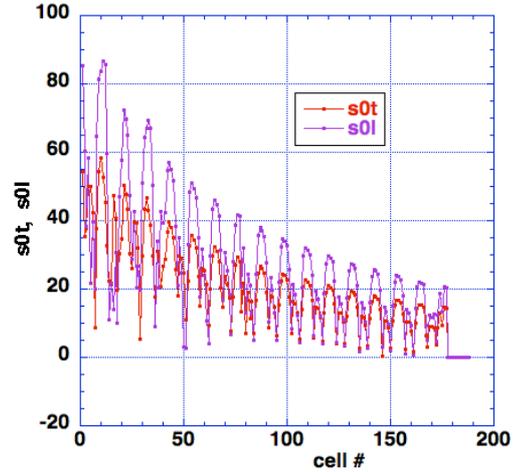

A 20-200MeV, 972MHz muon APF linac was designed using a linear *peratten* rule. The real transverse emittance input from the 0.034-20 MeV output (where β=0.542) = 0.00028. The dphi output ~±22° is multiplied by 3 = ±66 at 972 MHz, dw=0.2 MeV. The linac has 188 cells, length ~47 m. Iterations with the final one on both phis's and gapobl's resulted in 100% transmission and acceleration with no aperture restriction, and 55% with a constant aperture of 1 cm. Fig. 19 shows the phase advance and acceptances.

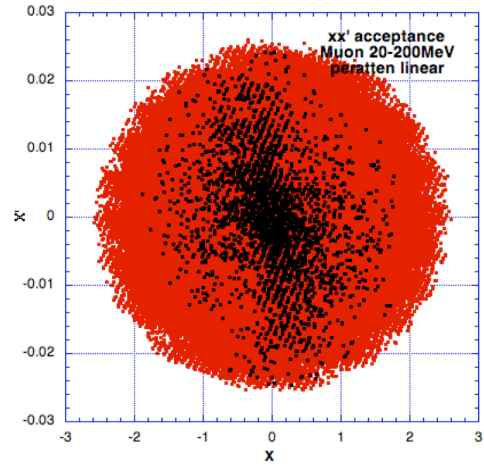

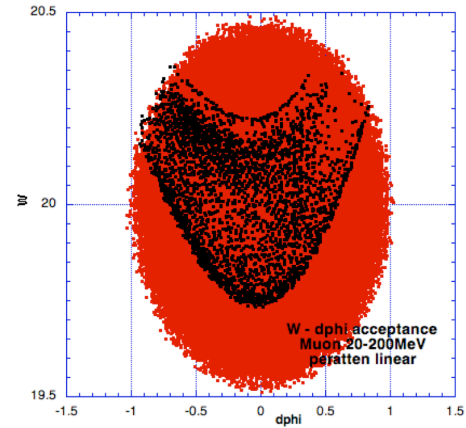

*Fig. 19. a. Zero-current phase advances for muon 20-200 MeV, 972MHz APF linac. b. & c. Transverse and longitudinal acceptances.*

A 0.340-200MeV, 972MHz muon APF linac, 251 cells, ~54 m, resulted in ~65% transmission and acceleration, Fig. 20.

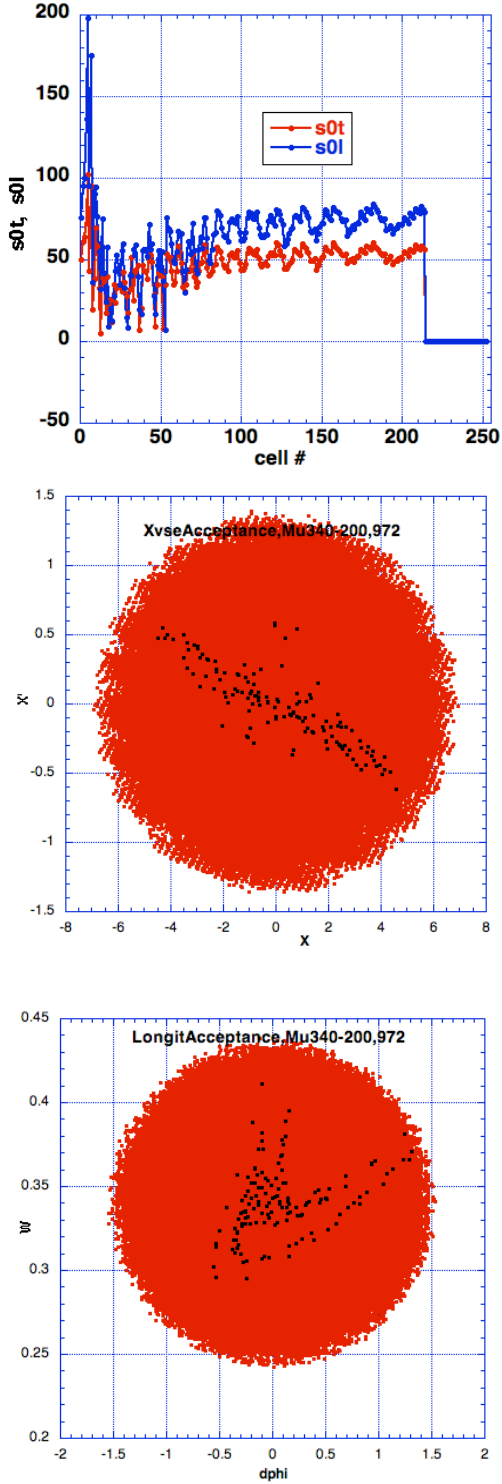

*Fig. 20. a. Zero-current phase advances for muon 0.340-200 MeV, 972MHz APF linac. b. & c. Transverse and longitudinal acceptances.*

Again it is seen that designs can be efficiently found and optimized for long sequences with large energy gain factors, upon which realistic practicalities can be investigated.

### VII. Model Enhancement

The simple but efficient tools now implemented in LINACSapf can readily be extended. Some steps would be to:
- Extend LINACSapf to have the option for a more accurate gap models, such as an analytical one, or a field map.
- Add space-charge. The preliminary design procedure would use the rms envelope method of LINACS, and afford "inside-out" design from desired beam space-charge physics considerations, including the possibility of using equipartitioning if desired [18]. Final simulation could then incorporate standard space-charge routines.
- Investigate actual construction. The 'fish filet' structure suggested in [15] seems an ideal candidate.
- The above examples were designed with an average longitudinal electric field Eo = (1.6 to 1.8)*Kilpatrick factor at the designated frequencies. Thus they represent approximately the maximum strength fields that might be obtained throughout the linac. Many practical aspects, such as actual design of the structure and determination of possible peak surface fields, other phase advance strategies, other sequence and optimization strategies, etc. can now be investigated.
- APF can be used in addition to conventional magnetic focusing, and could be useful in minimizing the amount of additional magnetic focusing needed to handle the desired amount of beam current.

### VIII. Conclusions

APF linacs can now be another practical approach in the linac designer's repertoire, and can be considered as a candidate for any application, either alone or in conjunction with magnetic focusing.

Working tools for the generation and beam dynamics simulation of APF linacs using a general synchronous phase sequence have been developed in the program LINACSapf. A driver framework allows quick searching over the parameter space of an 7-parameter general APF phase sequence, to look for desired characteristics such as maximum transmission, maximum accelerated beam fraction, low transverse emittance growth, low output phase of energy spread, etc.

Optimization from the parameter mesh search result is then done with a constrained nonlinear optimization program, using essential accelerator physics information about the period of the smooth approximation phase advance for convergence and efficiency.

The program was used to demonstrate that very long APF linacs, with high energy gain factors, are possible, for the first time. Examples for zero-current muon and H+ APF linacs were demonstrated. Future steps to incorporate space-charge and more accurate elements are straight-forward.

## IX. Acknowledgements